\documentclass[12pt]{iopart}
\usepackage{amsmath}
\usepackage{graphicx}

\begin{document}

\title[Models of fluidized granular materials]
{Models of fluidized granular materials: examples
of non-equilibrium stationary states}

\author{Andrea Puglisi\dag\  Fabio Cecconi\ddag\ and Angelo Vulpiani\P}

\address{\dag\ Laboratoire de Physique Theorique
Batiment 210, Universit\`e de Paris-Sud, 91405 Orsay Cedex, France}

\address{\ddag\ 
INFM Center for Statistical Mechanics and Complexity and 
Dipartimento di Fisica Universit\`a ``La Sapienza'' 
Piazzale A.~Moro 2 I-00185 Rome Italy.}

\address{\P Dipartimento di Fisica Universit\`a ``La Sapienza''\\ 
INFM Center for Statistical Mechanics and Complexity (SMC) \\
INFN Sezione di Roma-1 ``La Sapienza'' \\  
P.le A.~Moro 2, I-00185, Rome Italy.}

\begin{abstract} 
We review some models of granular materials fluidized by means
of external forces such as, random homogeneous forcing with damping,
vibrating plates, flow in an inclined channel and 
flow in a double well potential. All these systems show the presence of
density correlations and non-Gaussian velocity distributions. These
models are useful to understand the role of a kinetically defined
``temperature'' (in this case the so-called {\em granular temperature})
in a non-equilibrium stationary state. In the homogeneously randomly
driven gas the granular temperature is {\em different} from that of
the driving bath. Moreover two different granular materials mixed
together may stay in a stationary state with different
temperatures. At the same time, granular temperature determines (as in
equilibrium systems) the escape time in a double well
potential. 
\end{abstract}




\maketitle

\section{Introduction}
Granular materials such as sand, grains and powders exhibit a variety of
remarkable behaviors which, in the last decades, 
have been extensively through a number of experiments, computer 
simulations and analytical techniques~\cite{nagel,po1,po2}. 
This paper aims to review the conceptual and technical
difficulties encountered when
the same statistical approach successfully applied to simple 
fluids is generalized and extended to study granular systems.
The question whether the dynamics of a 
collection of inelastic particles is amenable to an
hydrodynamical and even ``thermodynamical'' description is a longstanding and
still controversial issue of the general theory of granular matter.    

In 1995, Du, Li and Kadanoff~\cite{du95} proposed and studied a minimal
model of a one-dimensional granular gas where $N$ hard rods, 
constrained to move on the segment $[0,L]$,
interact by instantaneous binary inelastic collisions with a restitution
coefficient $r<1$. A thermal wall of temperature $T_b$, at the boundary 
$x=0$, 
prevents the system from the cooling caused by inelasticity.  
When the leftmost
particle bounces against the wall, it is reflected with a velocity drawn 
from a Gaussian distribution with variance $T_b$ and transfers the energy to 
the rest of the system. 
The main finding of the authors was that even at
very small dissipation $1-r \sim 0$, hydrodynamic equations failed to 
reproduce the essential features of simulations.
Simulations, indeed, showed that the
system sets onto an ``extraordinary'' state with most of the particles
moving slowly and very near the right wall, while most of the kinetic
energy is concentrated in the leftmost particle.  
Reducing the dissipativity
$1-r$ at fixed $N$, the cluster near the wall becomes smaller and smaller.
The authors also pointed out that a qualitative explanation of this clustering
phenomena could be found in the Boltzmann equation in the limit $N \to
\infty$, $r \to 1$ with $N(1-r) \sim 1$. 
We have reproduced the results of Du et al. and found that 
the breakdown of hydrodynamic approach can be ascribed to the 
peculiarities of the model.

First, the one-dimensional character represents generally an obstacle
to the development of the hydrodynamic theory even for elastic systems
since transport coefficients usually
diverge with system size at low dimension. Of course exceptions exist as
shown in Ref.~\cite{sela}, where, under some particular circumstances,
the hydrodynamics of a 1-d inelastic system has been worked out.

Second, this model lacks proper thermodynamic limit because
when $N,L\to\infty$ (with $N/L=\mbox{const}$), both the mean kinetic energy
and the mean dissipated power reduce to zero. 
This is consistent with the scenario 
suggested by the authors in which energy equipartition is broken and the
description of the system in terms of macroscopic smooth quantities no
longer holds. 
The proposed mechanism of energy injection may become strongly inefficient 
because, even at moderate inelasicities, it may involve only just 
few particles near the thermal wall. In this condition, thermodynamic observables such as 
mean kinetic energy or mean dissipated power are non-extensive quantities.

Third, the system has no proper elastic limit, indeed when the
dissipation is removed by setting $r\to 1$, the kinetic energy
increases indefinitely due to the mechanical action of the wall that
continuously injects energy into the system. 
The situations gets even worst upon reducing the energy injection to zero
to take the elastic limit: 
elastic collisions simply exchange velocities and the
initial velocity distribution does not evolve at all.
 
In the following, we present and study a class of models where the 
aforementioned ``pathologies'' are partly removed. 
The common feature of these models is the presence of a stochastic 
external driving which, acting statistically on each particle, 
destroys the anomalous configurations observed in Ref.~\cite{du95}.
Furthermore the action of a damping term guarantees the existence of 
a smooth elastic limit. 
We shall see that although such systems display a ``less pathologic'' 
bevaviour, the existence of an hydrodynamical interpretation 
is still critical and dubious because it is affected by general 
conceptual problems~\cite{kadanoff99}.
The same difficulties, on the other hand, are also encountered in the 
interpretation of experimental studies on forced granular systems.    
For instance, a well known experiment by Jaeger  
et al.~\cite{jaeger94} consider a container full of sand shaken from the 
bottom plate.
When the shaking is very rapid, observations indicate that a 
few-grain thick boundary layer forms near the floor. 
Particles inside this
layer move very quickly with sudden changes in their dynamics. 
At the top of
the container, instead, particles move ballistically undergoing very few
collisions in their trajectory (Knudsen regime). 
Both layers cannot be described by hydrodynamics 
because the assumption of smooth variation of the velocity field is not 
satisfied. The same happens in molecular gases, however, for granular
systems
such boundary layers are macroscopic, and this 
seriously affects the prediction capability of hydrodynamic theories.  
Furthermore, the lack of a neat scale separation between the mean free time 
and the vibration period makes any hydrodynamic approach practically 
meaningless. On the other hand, it is also intrinsically unable to 
describe the slow grain dynamics at slow tapping rates.  
In this case, indeed, the system reaches a sort
of mechanical equilibrium characterized by an almost absence of
motion~\cite{knight95}.  Such an equilibrium is reached at different
densities which - as the tapping goes on - slowly
change with ``history'' dependent evolutions.  This memory
effect can not be captured by the set of partial differential 
equations concerning ordinary hydrodynamics.
 
The paper is organized as follows. In section II, we describe a model
of granular gas introduced to remove the ``pathologies'' affecting
previous models, such as the lack of a well defined thermodynamic
or elastic limit.  In section III and IV, we discuss
simulations on different models for driven granular gas with non
homogeneous energy sources. In section V, we discuss
the fundamental problem of scale separation which undermines the
development of a general hydrodynamic theory for granular flows. 
After having summarized the main failures of thermodynamic and 
hydrodynamic approaches to granular systems, we present, in section VI, 
a numerical experiment where instead thermodynamic concepts positively
apply. In section VII we finally draw some conclusions.

\section{Homogeneous driving by random forces \label{sec:driven}}

In papers~\cite{puglisi98,puglisi99} some of us introduced a kinetic model to
describe a granular gas kept in a stationary state under the effect of both a
damping term and external stochastic forcing. This model aims to reproduce the
experimental situations in which an inelastic system is forced by 
shearing, shaking, air fluidization, and so on. 
All these energy sources supply the system with an ``internal energy'' able to 
randomize the relative particle velocities.
They basically act as a temperature source \cite{driven1,driven2} 
which favours the onset of steady regimes, but  
at the same time, introduce a systematic (non random) friction which can 
be modelled by an effective damping term in the particle dynamics.

The randomly driven granular gas defined in~\cite{puglisi98} consists of
an assembly of $N$ identical hard objects (spheres, disks or rods) of
mass $m$ and diameter $\sigma$. 
We shall set $m=1$ and $k_B=1$ (Boltzmann's constant) in the following
and assume that grains move in a box of volume $V=L^d$.
The dynamics of the system is the outcome of three
physical effects: friction with the surroundings, random accelerations
due to external driving, inelastic collisions among the grains.  The
first two ingredients are modeled in the shape of Kramers' equations  
between two consecutive collisions 
\begin{eqnarray}
\frac{d}{d t}{\mathbf x}_i(t) & =  {\mathbf v}_i(t)      \label{langev1}\\
\frac{d}{d t}{\mathbf v}_i(t) & =  -\frac{{\mathbf v}_i(t)}{\tau_b}
+ \sqrt{\frac{2T_b}{\tau_b}}\mbox{\boldmath $\eta$}_i(t) \label{langev2}
\end{eqnarray}
Parameters $\tau_b$ (decorrelation time) and $T_b$ (temperature)
characterize the properties of the external bath.  The function
$\mbox{\boldmath $\eta$}_i(t)$ is the standard white noise: $\langle
\mbox{\boldmath $\eta$}_i(t) \rangle = 0$ and $\langle
\eta^\alpha_i(t)\eta^\beta_j(t') \rangle =
\delta(t-t')\delta_{ij}\delta_{\alpha \beta}$ ($\alpha,\beta = x,y,z$).  
This
choice guarantees that Einstein's relation \cite{kubo78} is fulfilled 
in the elastic or collisionless regime. 
The inelastic collisions, instead, are considered at the kinetic level, 
because an impact instantaneously transform the velocities of the  
grains involved. When particles $i$ and $j$ collide, their velocities are 
instantaneously changed into new velocities according to the following 
collision rule:
\begin{eqnarray}
\label{collision_rule}
\mathbf{v}_i' &=\mathbf{v}_i-\frac{1+r}{2}((\mathbf{v}_i-\mathbf{v}_j)\cdot 
\hat{\mathbf{n}}) \hat{\mathbf{n}}  \label{collis1} \\
\mathbf{v}_j' &=\mathbf{v}_j+\frac{1+r}{2}((\mathbf{v}_i-\mathbf{v}_j)\cdot 
\hat{\mathbf{n}}) \hat{\mathbf{n}}  \label{collis2}
\end{eqnarray}
where $\hat{\mathbf{n}}$ is the unit vector along the direction joining 
the centers of the particles, $r$ is called the normal restitution 
coefficient.
These rules reduce the longitudinal component of the relative velocity 
for $0 \le r<1$, which instead is only inverted at $r=1$. 

This model has been studied in detail~\cite{puglisi98,puglisi99}
through simulations using Direct Simulation Monte Carlo
(DSMC)~\cite{bird} and Molecular Dynamics (MD) algorithms \cite{CeccoJCP}
as well. 
The first method treats collisions stochastically, assuming 
the hypothesis of molecular chaos between 
particles at a distance smaller than $\sigma_B$ (a parameter which is chosen 
to be smaller than mean free path). It can be regarded as a sort of spatially
inhomogeneous Monte Carlo technique. The second method instead implements 
the dynamics of the model without any approximation requiring, however,
much more computational resources.

In the dynamics of the $N$ particles defined by
Eqs. (\ref{langev1},\ref{langev2}) and (\ref{collis1},\ref{collis2}), the relevant
parameters are: the coefficient of normal restitution $r$, which
determines the degree of inelasticity and the ratio $\tau_b/\tau_c$ 
of the forcing characteristic time (bath) to the
``global'' mean free time between consecutive collisions.
On the basis of these two
parameters, the dynamics of our model exhibits two fundamental regimes: 

\begin{itemize}
\item a stationary ``collisionless'' regime occurring when $\tau_b \ll
\tau_c$. In this regime we expect that, after a transient time of order
$\tau_b$, the system reaches the stationary behaviour of
independent Brownian particles characterized by homogeneous density,
Maxwell velocity distributions and absence of correlations.
\item A stationary ``colliding'' regime obtained when $\tau_b \ge \tau_c$.
If collisions are inelastic, this condition corresponds to the
{\em cooling} limit, and for times larger than $\tau_b$, 
the model evolves with anomalous statistical properties. 
\end{itemize}

\begin{figure}
\begin{center}
\includegraphics[width=6.0cm,angle=-90]{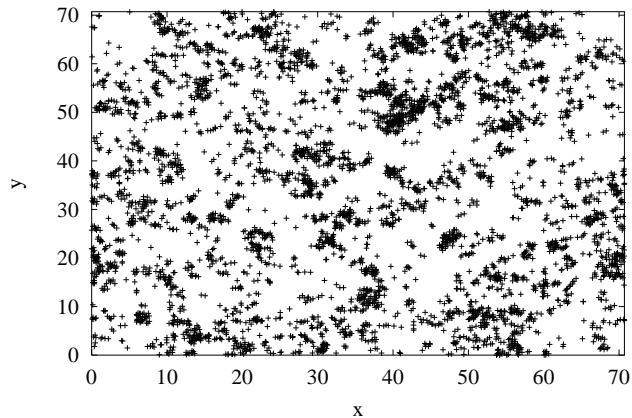}
\caption{Density snapshots in the homogeneously driven granular gas:
instantaneous plot of positions in 2d, in the inelastic
regime. $N=5000$, $\tau_c=0.5$, $\tau_b=100$ and $r=0.1$.}
\label{h_density1}
\end{center}
\end{figure}

Numerical simulations show that the thermodynamic limit on this model
is well defined, thus one of the problems affecting the Du et
al. system is solved.  The dynamics in the colliding
regime ($\tau_b \ge \tau_c$) and in the presence of inelasticity
($r<1$), results in a stationary state with a temperature $T_g$ always
lower than $T_b$. The granular temperatures approaches monotonically
$T_b$ as $r \to 1$, so that the elastic limit
can be safely taken without energy catastrophe. The fact that (when
$r<1$ and $\tau_b \ge \tau_c$) $T_g < T_b$ is the principal indication
that a our model of granular gas is a genuine non equilibrium system in a
statistically stationary state. This state is characterized by
inhomogeneous spatial arrangement of grains (clustering) and
non-Gaussian velocity distributions. Figure~\ref{h_density1} displays
a snapshot, from 2d simulations, of the particle positions in a strong
clustering regime.  The inelastic regime exhibits much stronger
density fluctuations than those occurring in the collisionless limit,
$\tau_b \ll \tau_c$, where, grains, instead, occupy the whole volume
uniformly with no correlations.  

\begin{figure}[htbp]
\includegraphics[width=5cm,angle=-90]{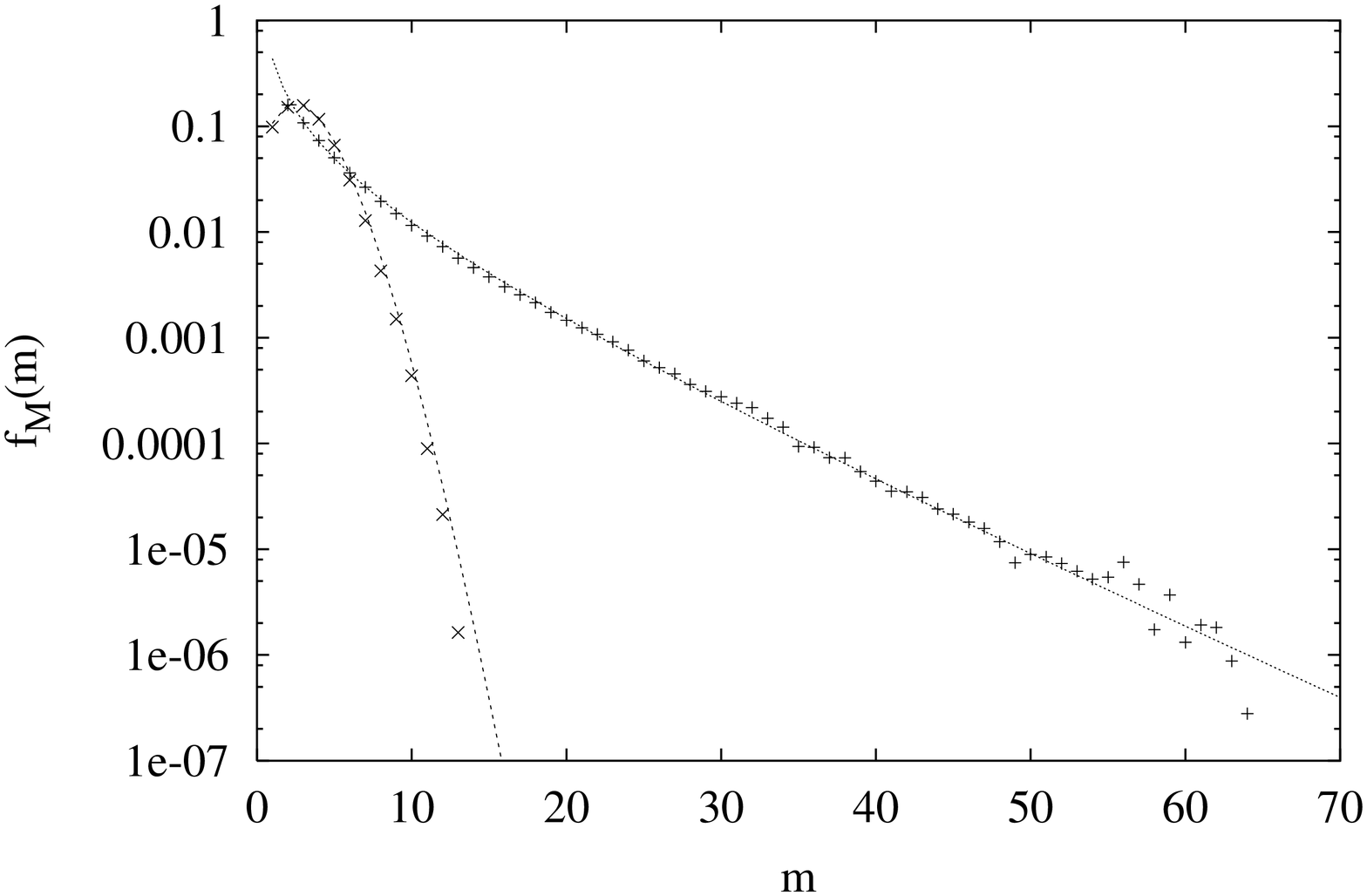}
\includegraphics[width=5cm,angle=-90]{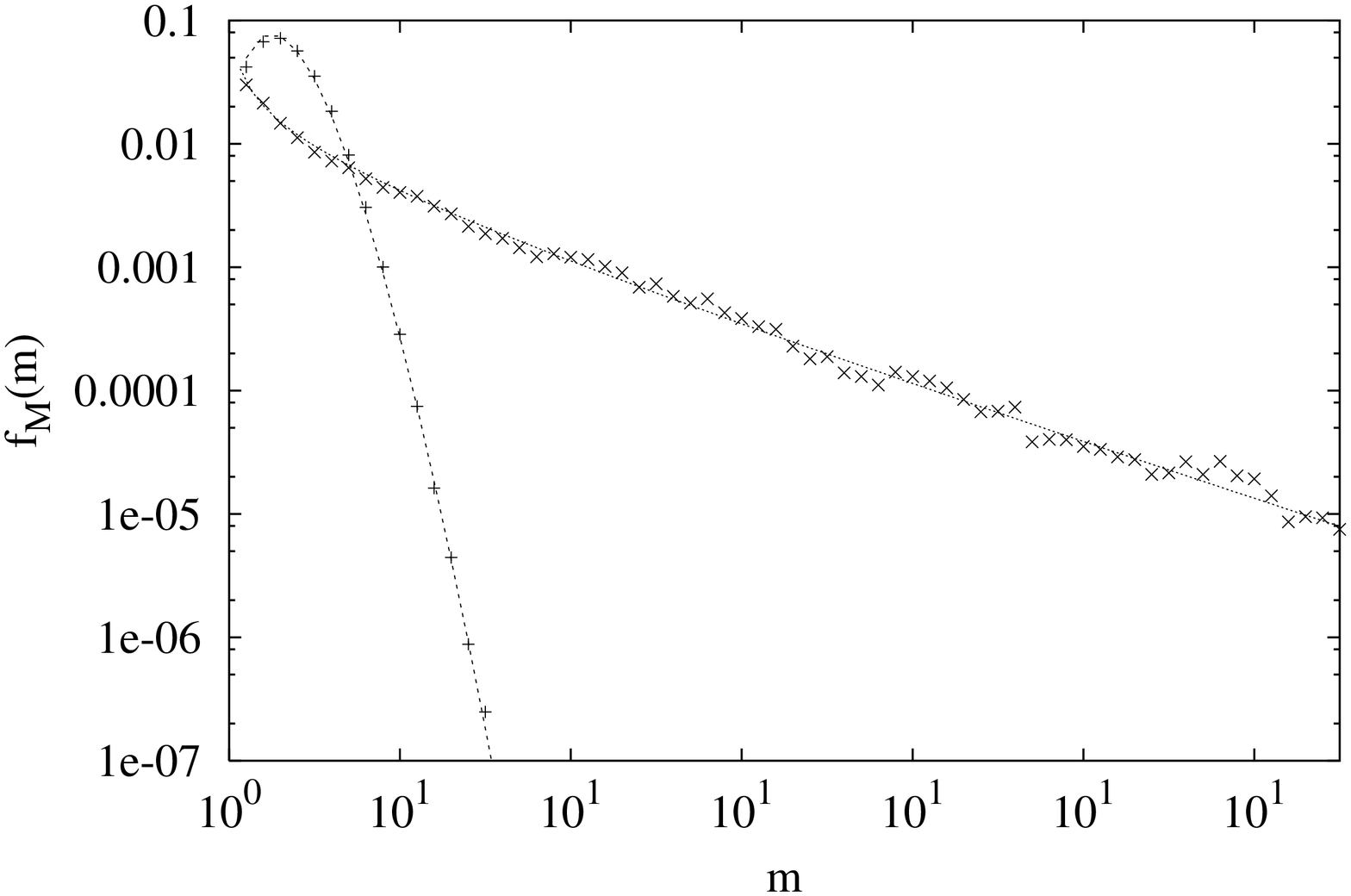}
\caption{Density fluctuations in the homogeneously driven granular
 gas. Left: 1d case, $N=300$, the $\times$ are obtained in a
 collisionless regime, while the $+$ corresponds to a colliding regime
 with $r=0.7$. Right: 2d case, with $N=500$ particles, the $+$ are
 obtained in an almost elastic collisionless regime, the $\times$
 corresponds to a colliding regime with $r=0.5$. In both figures the
 collisionless case (equal to an elastic case) is fitted by a Poisson
 distribution, while the colliding inelastic case is fitted by an
 inverse power law with exponential cut-off, as discussed in the
 text.}
\label{h_density2}
\end{figure}

In figure~\ref{h_density2}, we show the probability distribution of the 
``cluster mass'', $m$, defined as the number of particles found in a 
box of volume $V/M$. We divided the container of the system
into $M$ identical boxes with an elementary volume $V/M$.
In the collisionless regime, 
the number of particles in a box of size $V/M$ follows a 
Poisson-like distribution with average $\langle m_M \rangle = N/M$.
On the contrary, the colliding regime ($\tau_b \gg \tau_c$) generates 
very different density distributions which  
can be fitted by a power-law 
$m_M^{-\alpha_{cl}}\exp(-c_{cl}m_M)$ 
corrected by an exponential cut-off only due to finite size effects.  
In most of the simulations, we found $\alpha_{cl}>1$
and $1/c_{cl}$ slightly greater than $N/M$. 
The power law behaviour is the signature of self-similarity in the
distribution of clusters occurring with no characteristic size. 
These anomalous density fluctuations 
are not an artifact produced by the simulation technique
because they have been observed by
using both DSMC and MD algorithms. 
We have characterized the 
emergence of spatial correlations through the measure of the correlation
dimension $d_2$ (Grassberger and Procaccia~\cite{grassberger83}).
The latter is defined by 
scaling behaviour $C(R)\sim R^{d_2}$ of 
the cumulated particle-particle correlation
function
\begin{equation}
\label{correlation_function}
C(R) = \frac{1}{N(N-1)}\sum_{i \neq j} 
\overline{\Theta(R-|{\bf x}_i(t)-{\bf x}_j(t)|)} \sim R^{d_2}
\end{equation}
where the over-bar indicates the time averaging taken 
after the system reaches a steady regime, $R$ is the spatial resolution and
$\Theta(u)$ is the unitary step function.
\begin{figure}[htbp]
\begin{center}
\includegraphics[width=5cm,angle=-90]{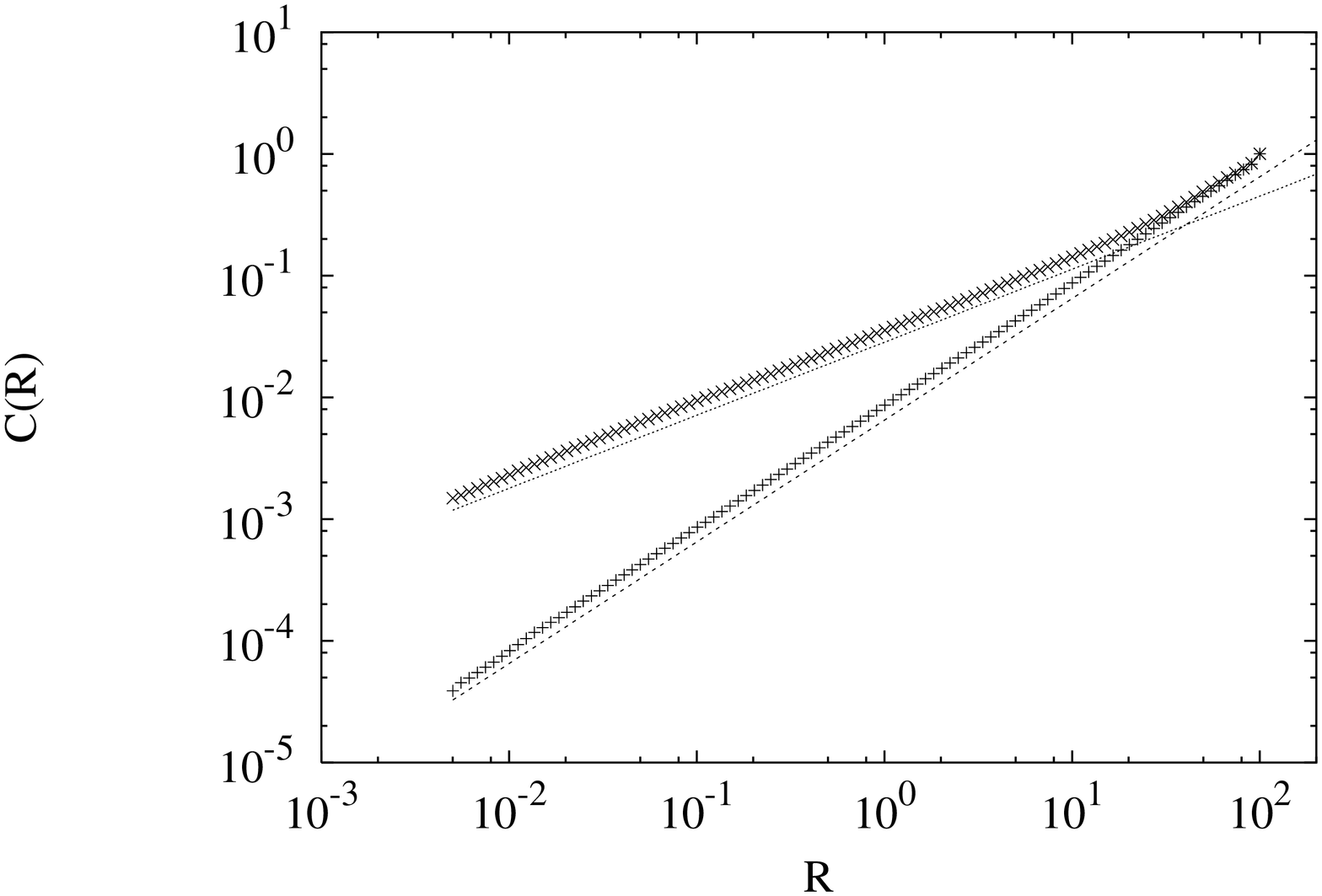}
\includegraphics[width=5cm,angle=-90]{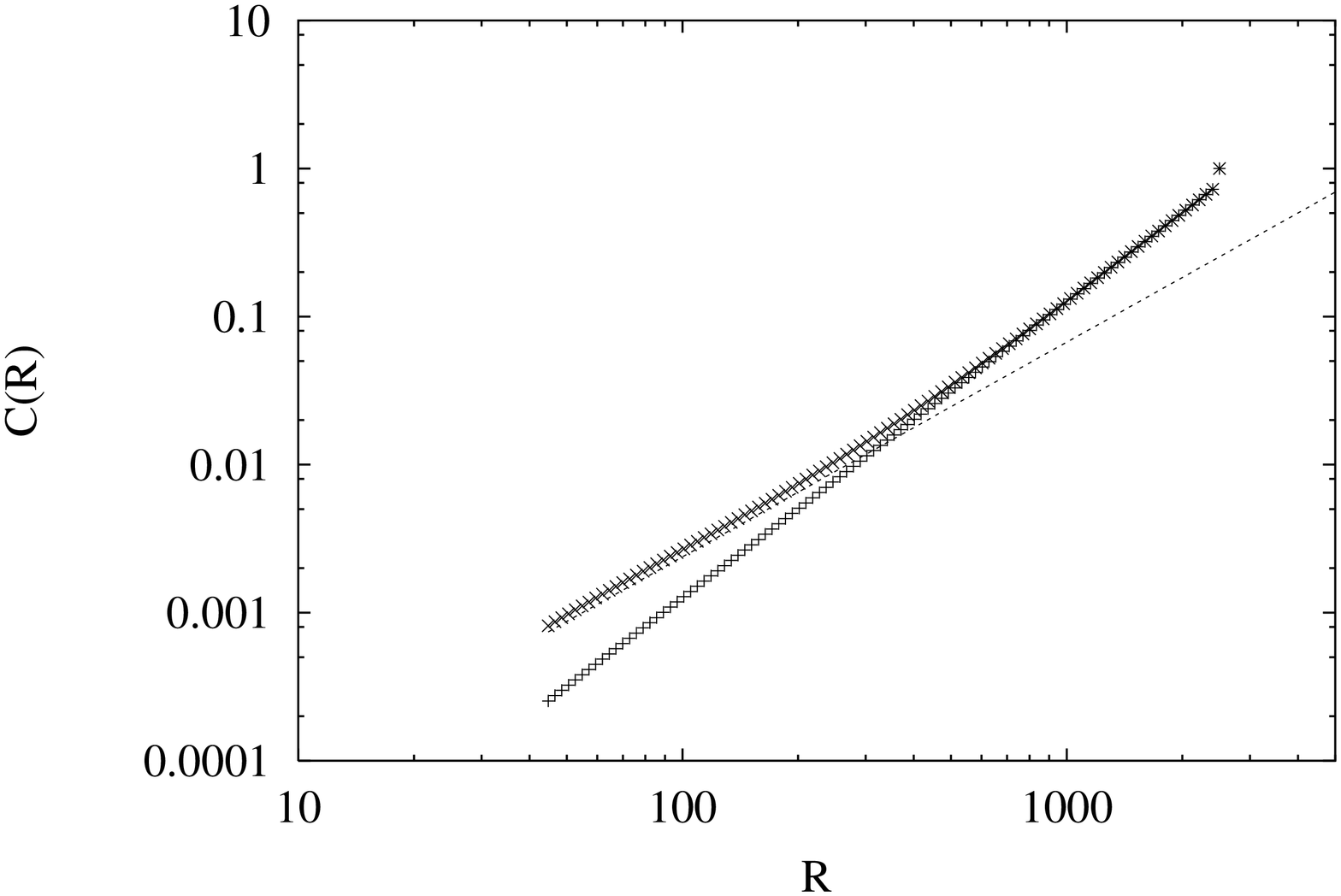}
\caption{Density-density correlation function in the homogeneously
driven granular gas. Left, in 1d ($N=2000$) and, right, in 2d
($N=5000$). In both graph the top (lower slope) curve corresponds to a
colliding inelastic case, while the curve with the larger slope 
(corresponding to the exact dimensionality of the space) is obtained in a 
collisionless regime ($\tau_c \gg \tau_b$).}\label{h_C}
\end{center}
\end{figure}
For homogeneous density, $d_2$ coincides with the euclidean 
dimension $d_2=d$, while the result $d_2<d$ is an indication of a 
{\em fractal} (self-similar) density. 
Model simulations carried out in the collisionless regime lead always 
to homogeneous distributions of particle (Fig.~\ref{h_C}),
while fractal densities often occur
in inelastic colliding regimes ($\tau_b \gg \tau_c$). 
This is consistent with the scenario provided by 
the mass distribution of clusters discussed above. 
\begin{figure}[htbp]
\begin{center}
\includegraphics[width=5cm,angle=-90]{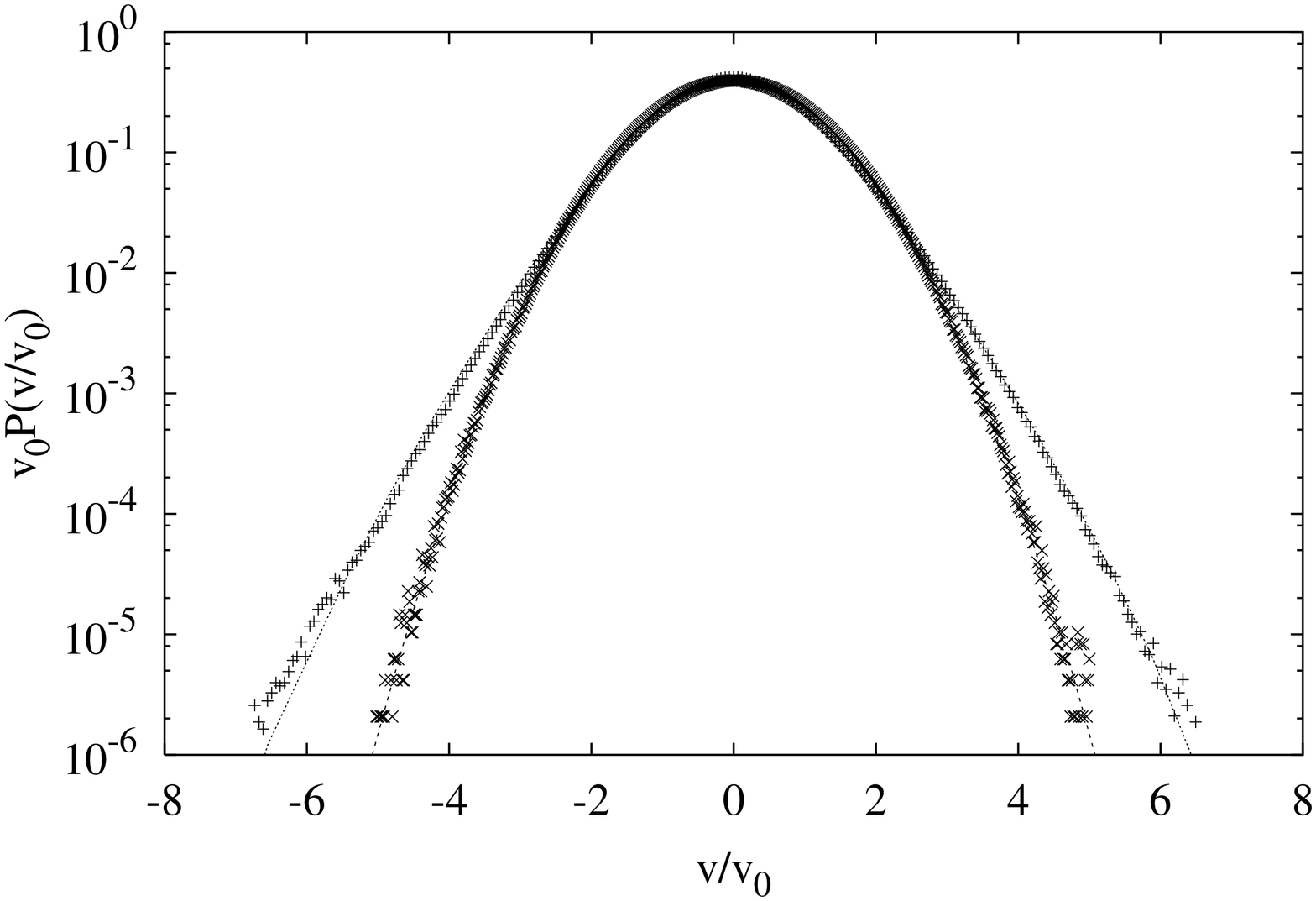}
\includegraphics[width=5cm,angle=-90]{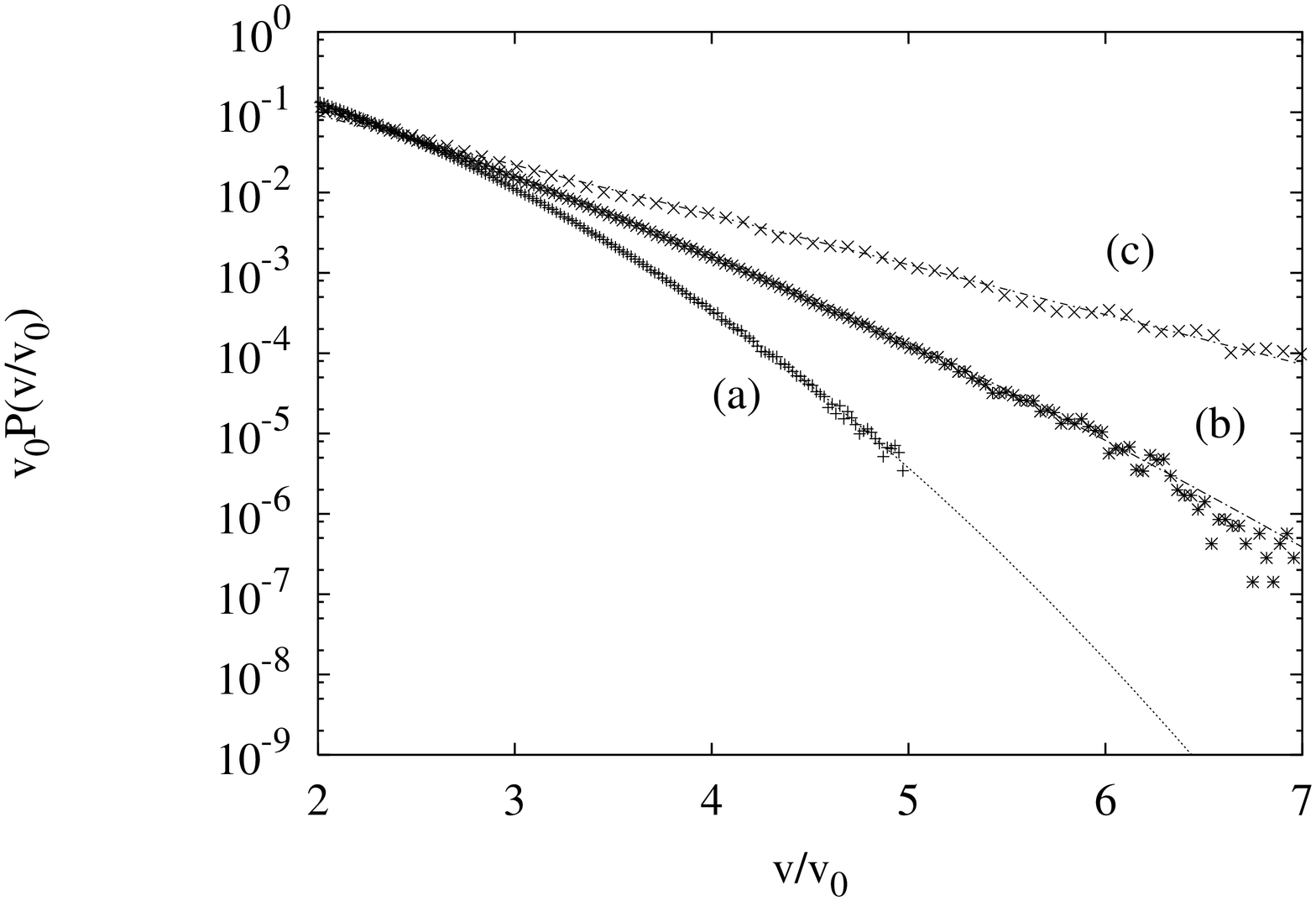}
\caption{Velocity fluctuations in the homogeneously driven granular
gas, in 1d (left) and 2d (right). The two situations in the left graph
correspond to a collisionless regime (Gaussian fit) and colliding
inelastic regime (non-Gaussian fit).  The three situations in the
right graph corresponds to: a) a collisionless regime ($\tau \ll
\tau_c$) with a distribution well fitted by a Gaussian; b) an
intermediate regime ($\tau_b \sim \tau_c$) with very low restitution
coefficient ($r=0.5$), fitted by an $\exp(-v^{3/2})$ curve; c) a
strongly colliding regime fitted by an exponential distribution.}
\label{h_v}
\end{center}
\end{figure}
Another peculiarity of a driven granular gas is the
behavior of the velocity distribution $P(\mathbf{v})$.
Typical $P(\mathbf{v})$ for our model in 1-d and 2-d 
simulations are shown in figure~\ref{h_v}.
We see a strong difference between the collisionless
(or elastic) regime, 
$\tau_c \gg \tau_b$ and the inelastic colliding regime
$\tau_c \ll \tau_b$. 
The collisionless regime is characterized by a Gaussian $P(\mathbf{v})$ 
while, in the colliding regime, a non-Gaussian behavior appears
as an enhancement of high-energy tails and the fitting procedure of
such tails provides the direct evaluation of the deviation from the
Gaussian regime.
In our simulations, we found the evidence for $\exp(-v^{3/2})$ tails, 
in agreement with the theoretical prediction by
Ernst and van Noije~\cite{noije98c}.
Remarkably, we see from the right panel of figure~\ref{h_v}, that our model
in the regime $\tau_c \ll \tau_c$, is able to reproduce also 
the exponential tails $\exp(-v)$ expected by
the theory of Ref.~\cite{noije98c} for ``homogeneous cooling states''.  
However it is worth noticing that the result
was derived with the assumption of spatial homogeneity, condition 
violated by our simulations when the system undergoes clustering.

To our knowledge, experimental measurements of velocity distributions
have been performed only recently and noticeably only for steady
state granular systems under some sort of energy injection. We recall
some of the used laboratory setups where non-Gaussian distribution have
been observed:
\begin{description}
\item[a]
Vibration of the bottom of a 3D granular system
(Losert et al.~\cite{losert99b})
\item[b]
Vertical vibration of an horizontal plate with a granular mono-layer on
the top of it (Olafsen and Urbach~\cite{olafsen98,olafsen99})
\item[c]
Vibration of the bottom side of an inclined plane with a very dilute
granular mono-layer rolling on it, under the presence of gravity
(Kudrolli and Henry~\cite{kudrolli00})
\item[d]
Vibration of the bottom of a granular system confined in a vertical plane 
(Rouyer and Menon~\cite{rouyer00})
\end{description}

\begin{figure}[htbp]
\begin{center}
\includegraphics[width=5cm,angle=-90]{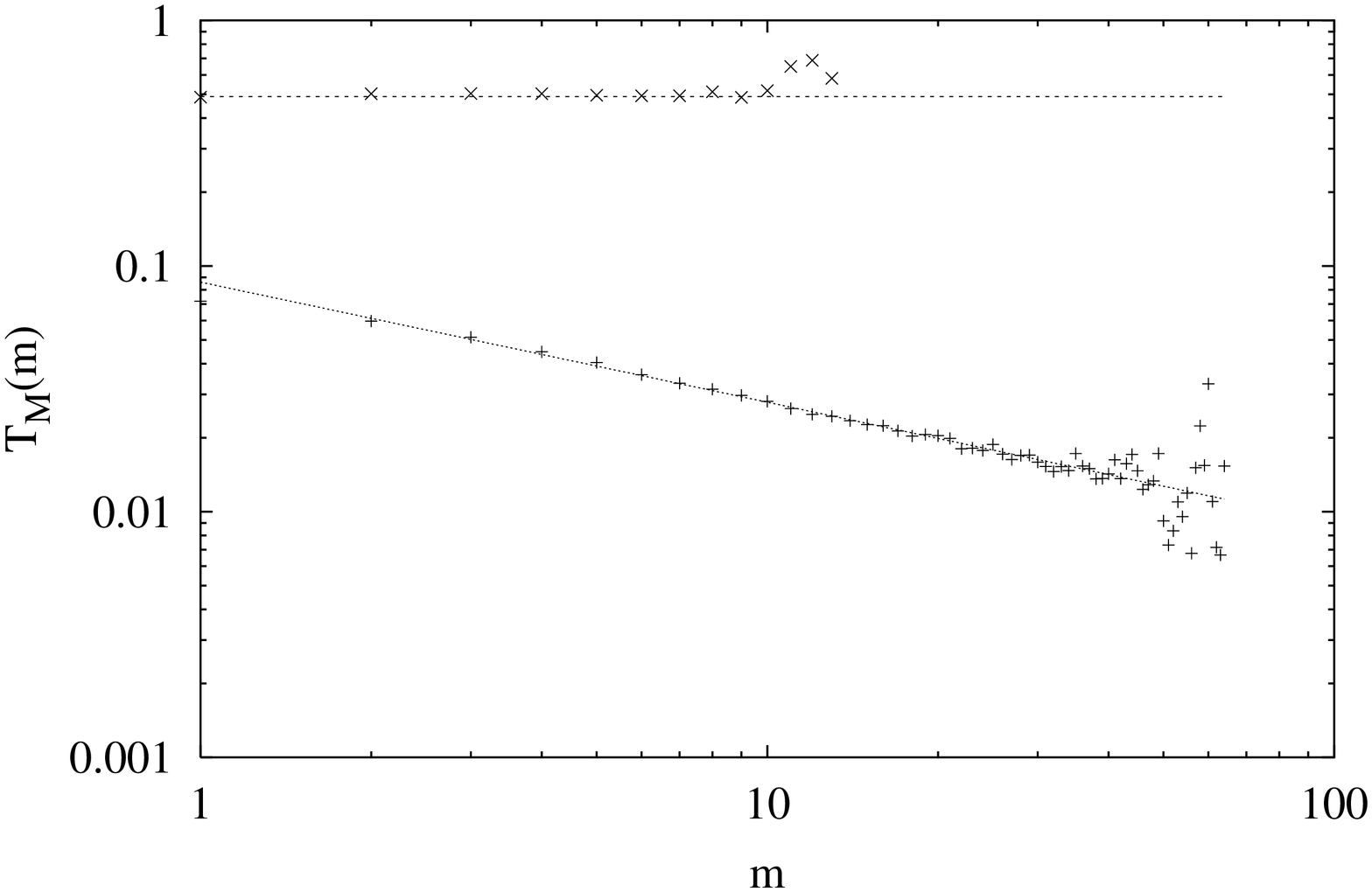}
\includegraphics[width=5cm,angle=-90]{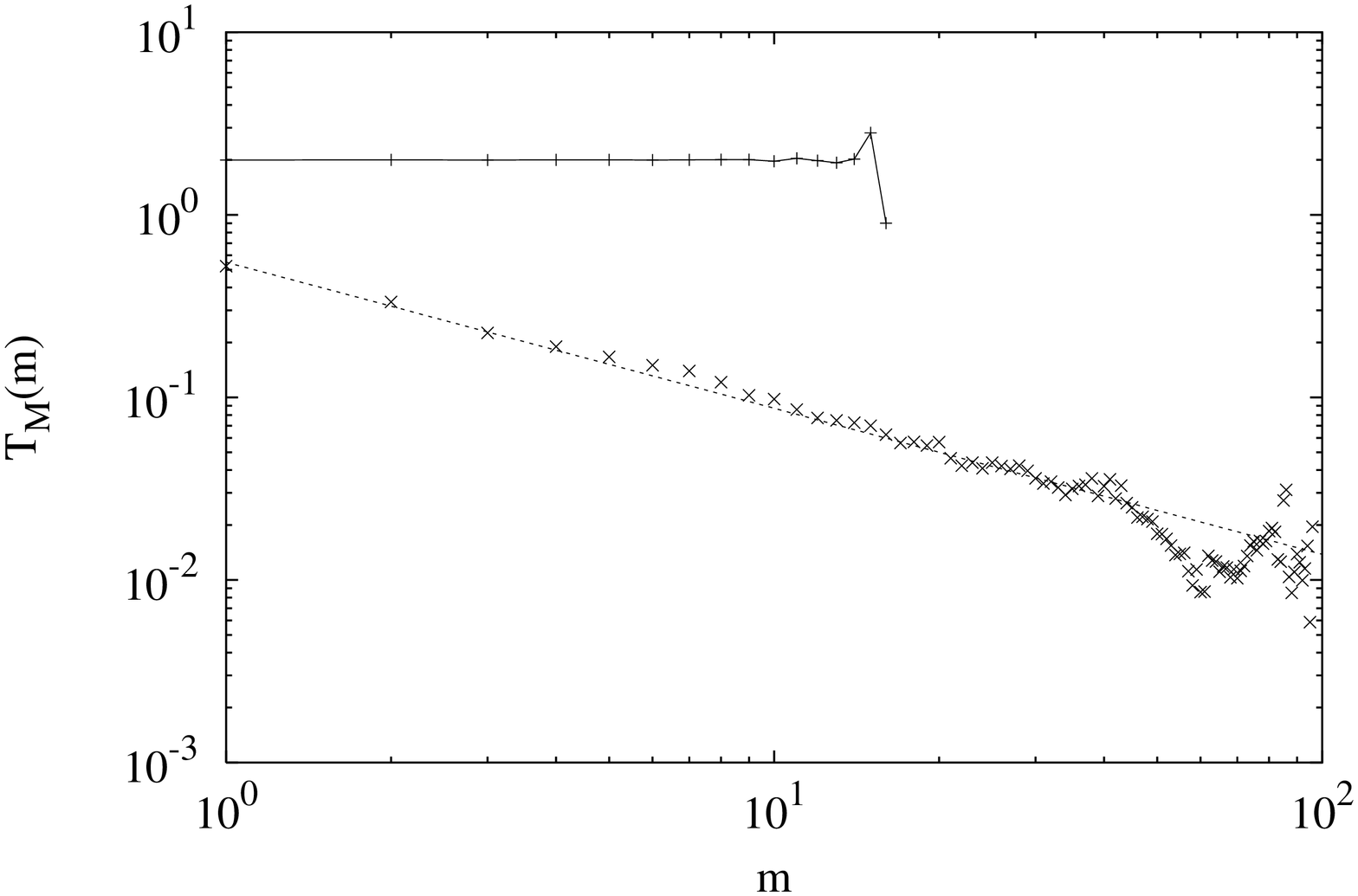}
\caption{Temperature fluctuations in the homogeneously driven granular
gas, in 1d (left) and 2d (right). In both figures the horizontal curves are
obtained in a collisionless (or quasi-elastic) system, while the
inverse power laws are observed in a colliding inelastic case.}\label{h_T}
\end{center}
\end{figure}

One of the arguments given by Goldhirsch~\cite{goldhirsch99} to explain
the existence of a general clustering instability starts from an heuristic
estimate of the local temperature (granular temperature $T_g$) as a function
of particle density $n$ and local shear rate
\begin{equation}
T_g \propto \l_0^2 \propto n^{-2},
\label{eq:tg_vs_n}
\end{equation}
$l_0$ is the particle mean free path. The above relationship 
remains meaningful at time scales shorter than the decay time of the 
shear rate.
Thus the local scalar pressure is supposed to decrease  
at larger densities $p = n T_g \propto n^{-1}$ implying an  
instability, because a positive fluctuation in the number of
particles, in a given region, causes a reduction in the local
pressure which attracts many other particles 
under the effect of pressure gradient.
However, formula~(\ref{eq:tg_vs_n}) strictly holds in the cooling regime 
and does not apply to our driven system for which  
the relation between local temperature $T_g$ and local density is
very different.
Simulations, in fact, indicate that the mean square velocity $T_g(k)$, 
in the $k$-th box, as a function of the
number of grains $m_k$ in that box exhibits a more general power-law
behavior (the total volume has been divided in $M$ identical boxes).
As expected, in the clustering regime, the distribution of the number of
particles in a box (cluster masses) presents a power-law decay with an
exponential cut-off. 
This induces also a non trivial power-law
behavior in the relation $T_g(k)$ versus $m_k$ as reported in
figure~\ref{h_T}, 
where we see that in the collisionless (or elastic) regime the
local temperature remains nearly constant so does not depend upon 
the cluster mass $m$. In the
inelastic condition, instead, the local temperature appears to be a power 
of the
cluster mass, $T_g(m) \sim m^{-\beta}$ with $0 < \beta < 1$. 
This relation ensures that the ``clustering catastrophe''
(particles falling in an inverted pressure region) 
can not occur because the scalar pressure $p = n T_g \propto
n^{1-\beta}$ increases with the density since $1-\beta$ is 
a positive exponent.  
Moreover, by using the previous result on the fractal correlation
dimension $d_2$ (Eq.~(\ref{correlation_function})), 
we can give an estimate of the length-scale dependence of
the temperature. In fact, if we assume that the scaling relation for 
the temperature is valid at
different spatial scales, we can replace the {\em density} to the
{\em number of particles} in the expression for $T_g$, i.e.
$T_g(n) \sim n^{-\beta}$.
Since the local density is expected to follow the scaling
$n(R) \sim R^{-(d-d_2)}$, the local temperature
follows the law $T_g(R) \sim R^{\beta(d-d_2)}$ and accordingly 
the local pressure behaves like $p(R) \sim R^{-(1-\beta)(d-d_2)}$. 
In conclusion, the density and the 
pressure both decrease with the length scale $R$, while the temperature 
increases.
This scale dependence of the macroscopic fields is evidently at odds with the
possibility of separating mesoscopic from microscopic scales 
and therefore the hydrodynamical description can not be attempted.
The inability of granular temperature to play the same role of kinetic
temperature in equilibrium statistical physics (for example being
equal to the temperature of the thermostat in the stationary
asymptotic regime) is further demonstrated by models of granular
mixtures~\cite{mixtures_theo}. Two granular materials, fluidized by the same
kind of homogeneous random driving mechanism, show up different
kinetic temperatures in agreement with experiments~\cite{mixtures_exp}. 
However, in the last section of this article, we discuss a toy model
where granular temperature recovers a role similar to ``thermal temperature'' 
making the situation even more complex.

\section{Systems with a vibrating floor}
Recent experiments~\cite{kudrolli00} have investigated the effect of gravity 
on driven granular materials.
Gravity, as a  uniform force field, has no consequences on relative 
velocities and thus on the sequence of collisions. 
It simply accelerates the center of mass of a
granular gas and its action becomes relevant only when studied in the 
presence of particular boundary conditions that break the Galilean 
invariance (horizontal planes or plates).  
\begin{figure}[htbp]
\begin{center}
\includegraphics[width=4cm,angle=-90]{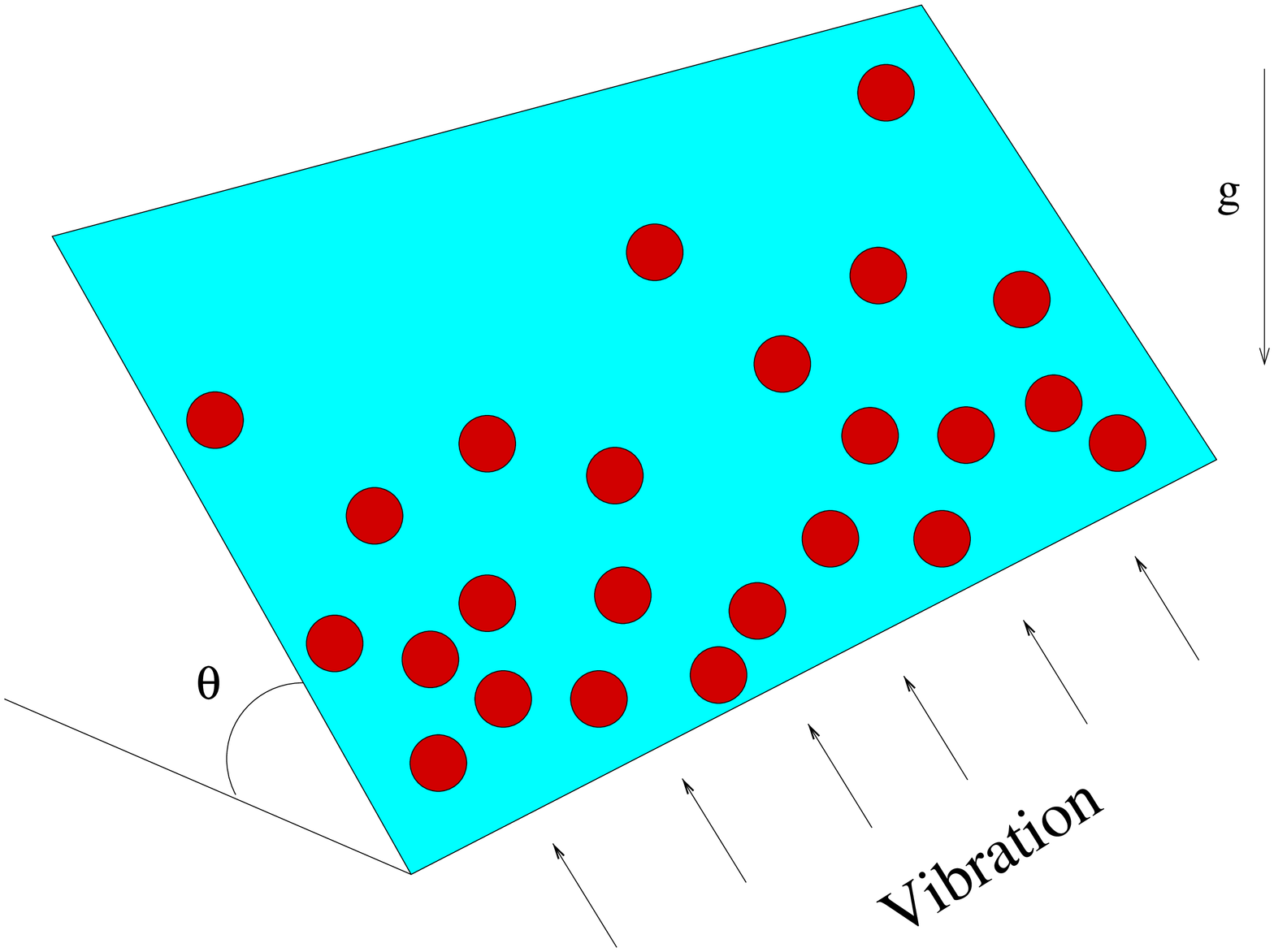}\hspace{1cm}
\includegraphics[width=4cm,angle=-90]{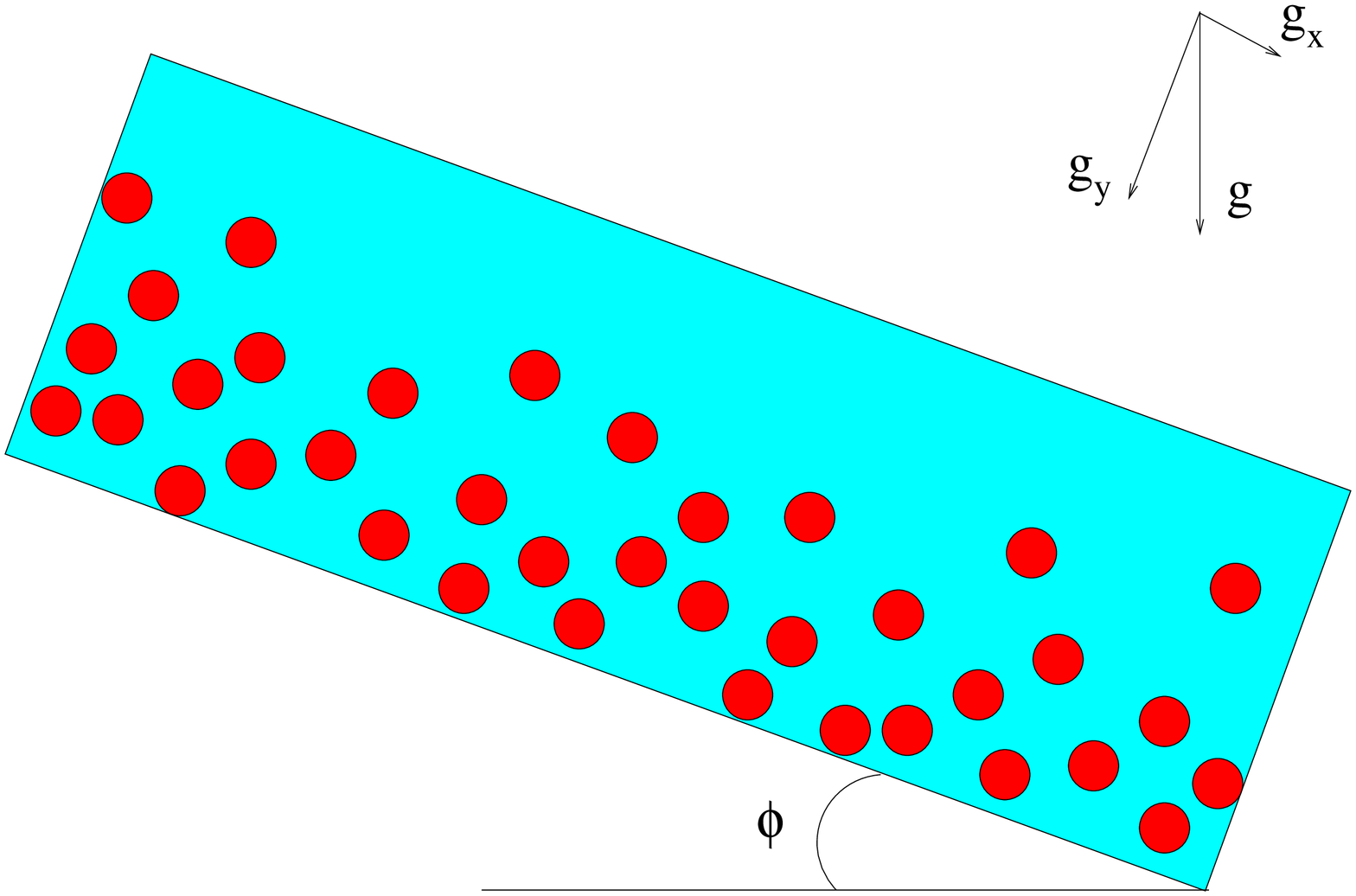}
\caption{Left: sketch of non-homogeneous model with gravity and
vibrating bottom. Right: sketch of non-homogeneous model with gravity
and inclined bottom.}
\label{i_sketch}
\end{center}
\end{figure}
A plate has an important role in disordering the velocity distributions 
especially when it vibrates
in the presence of gravity which, driving the grains toward the horizontal 
plates, makes the randomization process even much efficient.

The left frame of figure~\ref{i_sketch} sketches the geometry
setup of an experiment conducted in Ref.~\cite{kudrolli00} consisting
of a plane of size $L_x \times L_y$
inclined by an angle $\theta$ with respect to the horizontal.
The top and the bottom wall confine particles to move in such a plane 
under the action of an effective gravitational 
force $g_e=g\sin{\theta}$.
In our simulations, we reproduced the geometry and applied periodic 
boundary conditions in the horizontal direction. We assumed that both the
top and the bottom walls of the plane are inelastic with a restitution
coefficient $r_w$. The transfer of energy and momentum into the system
is realized, in our modeling through either
sinusoidal or stochastic (thermal) vertical vibrations of the
bottom wall. In the first case a
particle bouncing onto this wall is reflected with a vertical
velocity component: $v_y'=-r_wv_y+(1+r_w)V_w$, where $V_w=A_w
\omega_w \cos(\omega_w t)$ is the vibration velocity of the wall.
In the second case, a particle, after the 
collision against the wall, acquires randomly new velocity components 
$v_x \in (-\infty,+\infty)$ and $v_y\in (0,+\infty)$ 
with probability distributions 
$P(v_y)=\frac{v_y}{T_w}\exp(-\frac{v_y^2}{2 T_w})$
and
$P(v_x)=\frac{1}{\sqrt{2 \pi T_w}}\exp(-\frac{v_x^2}{2 T_w})$ respectively,
where $T_w = (A_w \omega_w)^2/2$ is the mean energy supplied
by the wall to the gas in a period of oscillation.

For moderate vibration intensities, the model sets onto an 
highly fluidized stationary phase which resembles turbulence. 
The time evolution of density and velocity fields exhibits 
an intermittent-like behavior characterized by rapid and large fluctuations, 
sudden explosions (bubbles) followed by the formation of  
large particle clusters traveling coherently downward under the
action of gravity.
\begin{figure}[htbp]
\begin{center}
\includegraphics[width=6.0cm,angle=-90]{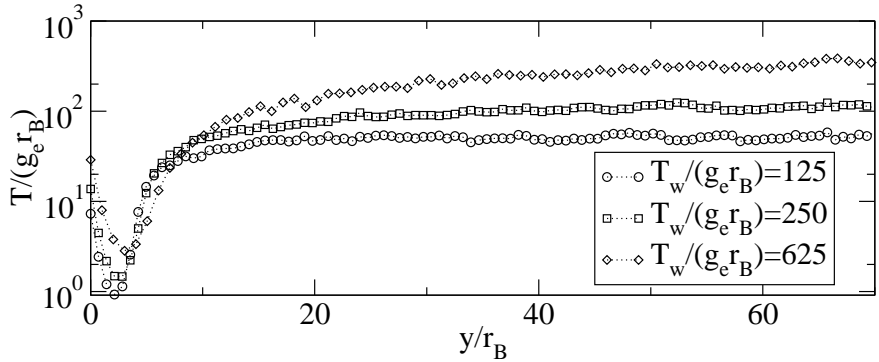}
\caption{Temperature profiles versus the rescaled height $y/r_{B}$
at three normalized forcing intensities $T_w/(g_e r_B)$,
for the model with gravity and periodic vibrating bottom
(Fig.~\ref{i_sketch})-left.}
\label{i1_T}
\end{center}
\end{figure}
In figure~\ref{i1_T}, we report the steady temperature profile
$T_g(y)$ for our system as obtained from simulations;
a minimum of $T_g(y)$ is clearly visible near y=0, position of the bottom
wall.
The parametric plot of granular temperature $T_g$ versus the particle 
density $n$ determines a power law $T_g \sim n^{-\beta}$ which closely
recalls the algebraic tails already observed in the behaviour of the
randomly driven model (sec. \ref{sec:driven}).

\begin{figure}[htbp]
\begin{center}
\includegraphics[width=5cm,angle=-90]{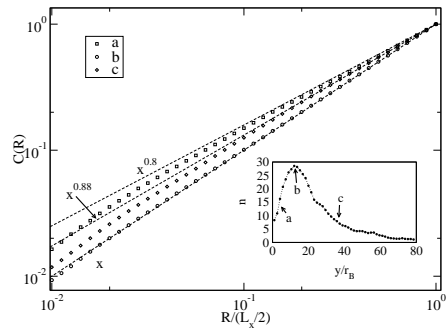}
\caption{Function $C_{\Delta y}(y,R)$ for three stripes at different heights.
The density profile as a function of the rescaled height is reported in the inset,
labels $a,b,c$ indicate the average densities and the heights of the
slabs chosen to compute the three correlation curves in the main plot.}
\label{i1_C}
\end{center}
\end{figure}

As before, the particle-particle correlation function is the useful
indicator to quantify the degree of spatial arrangement in the system.
In this case, the suitable quantity to measure is the particle-particle 
correlation function $C_{\Delta y}(y,R)$ conditioned to the height $y$, 
i.e. computed over the horizontal slab 
$B(y,\Delta y) = [y-\Delta y/2,y+\Delta y/2]\times [0,L_x]$.  
Data collected during simulations 
show a power law behavior $C_{\Delta y}(y,R) \sim
R^{d_2(y)}$ (Fig.~\ref{i1_C}). 
Again for homogeneous densities, $d_2$ is
expected to coincide with the topological dimension of the box 
$B(y, \Delta y)$, so we obtained $d_2 = 1$ for all the resolution 
$R\gg\Delta y$ at which the box
appears as a ``unidimensional'' stripe. While 
we found $d_2=2$,
at resolutions $R\ll\Delta y$, because the slab appears as two-dimensional
object.
When the regime of inelastic collisions is switched on,  
clustering processes, characterized by values of 
$d_2$ lower than the topological dimension, appear in some of the
analyzed stripes, as clearly seen by the three slopes of the log-log plot of
Fig.~\ref{i1_C}.
These three power-laws refers to three slabs at different heights
(labelled by $a$ $b$ and $c$) and very different density conditions
marked by the arrows in the inset showing the density profile.
When the density is not too high, the fit performed in the region
$R\gg \Delta y$ yields always an exponent smaller than $1$.

\begin{figure}[htbp]
\begin{center}
\parbox{5.8cm}{\includegraphics[width=5.8cm,clip=true]{Fig9a.ps}}\hspace{1cm}
\parbox{4.7cm}{\includegraphics[width=4.7cm,clip=true,angle=-90]{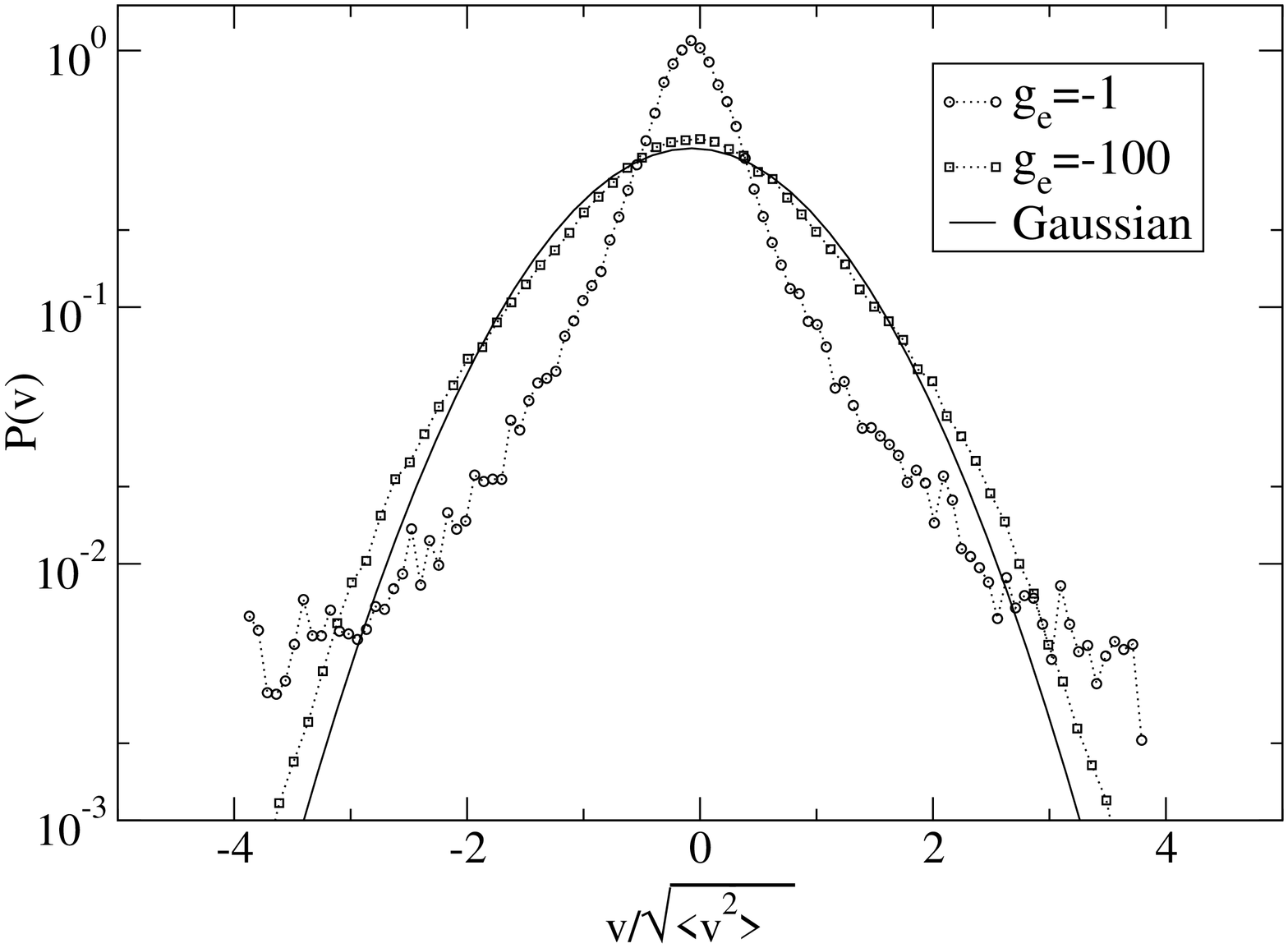}}
\caption{Left: collapse of horizontal velocity distributions in
slabs $a,b,c,d$ at different heights (see inset), for the model with
gravity and vibrating bottom.
The log-linear plot highlights the non-Gaussian character
of the tails.  
The inset displays the density profile as a function of the 
height from the bottom wall. Labels $a,b,c,d$ indicate the corresponding heights of the stripes 
chosen to sample the
velocity distributions. 
Right: horizontal velocity distributions for the same system with two different inclination 
of the plate, i.e. with two different values of the effective gravity $g_e$.}
\label{i1_v}
\end{center}
\end{figure}
In the left frame of Fig.~\ref{i1_v}, instead we report
the typical distributions of horizontal velocities for particles belonging
to stripes at different levels (densities) above the bottom wall 
($a,b,c,d$ in the inset). Axes variables are
properly rescaled to obtain a data collapse. Again, the distributions
appear to be
non-Gaussian and their broadening, namely the granular temperature
$T_g(y)$, is height dependent. The same behavior can be observed for
both periodic or stochastic vibrations. 
The right frame of Fig.~\ref{i1_v} indicates that the distribution of 
horizontal velocities becomes more and more Gaussian when the angle of 
inclination is risen. 
This trend toward a Gaussian behaviour, in perfect agreement with 
experimental observations \cite{kudrolli00}, is a consequence of 
a large inclination, which, enhancing the collision rate against the wall,  
favours the ``randomization'' of velocities.
According to the analogy between vibrating walls and heating baths,
this scenario is consistent with that 
observed for the randomly driven model where
larger ``heating rates'' (decrease in $\tau_b/\tau_c$)
determine a transition from non-Gaussian to Gaussian regime. 

\section{Acceleration onto an inclined plane}
In the context of non homogeneous driven granular gases, we analyzed a 
second more interesting model~~\cite{baldassarri01c} 
whose geometry is sketched in right frame of figure~\ref{i_sketch}.
The ``set-up'' consists of a two dimensional
channel of depth $L_y$ and length $L_x$, vertically confined by a
bottom and a top inelastic wall, periodic boundary conditions are applied
in the direction parallel to the flow. The channel is tilted up by an
angle $\phi$ with respect to the horizontal line, so gravity has both
components $g_x=g\sin{\phi}$ and $g_y=g\cos{\phi}$.
This model mimics the experiment performed by Azanza {\em et
al.}~\cite{azanza97}, where a stationary flow in a two-dimensional
inclined channel was observed at a point far from the source of the
granular material.
The assumption of periodic boundary conditions in
the flow direction is consistent with the observed
stationary regime reached upon the balance between gravity drift 
and damping effect due to inelastic collisions.

Simulated density, velocity and temperature profiles well
reproduce those measured in experiments~\cite{azanza97}. They, indeed,
show the existence of a critical height, $H \sim 6 \sigma_B$, 
corresponding to the separation between two different dynamical regimes. 
Below $H$, profiles look almost linear
especially the density and velocity ones, while above $H$ profiles rapidly
change and become nearly constant.
These changes in the behavior  
can be explained by the fact that below $H$, transport is mainly 
dominated by collisional exchange, while above $H$ it is mainly associated 
to ballistic flights. 
\begin{figure}[htbp]
\begin{center}
\includegraphics[width=5.0cm,clip=true,angle=-90]{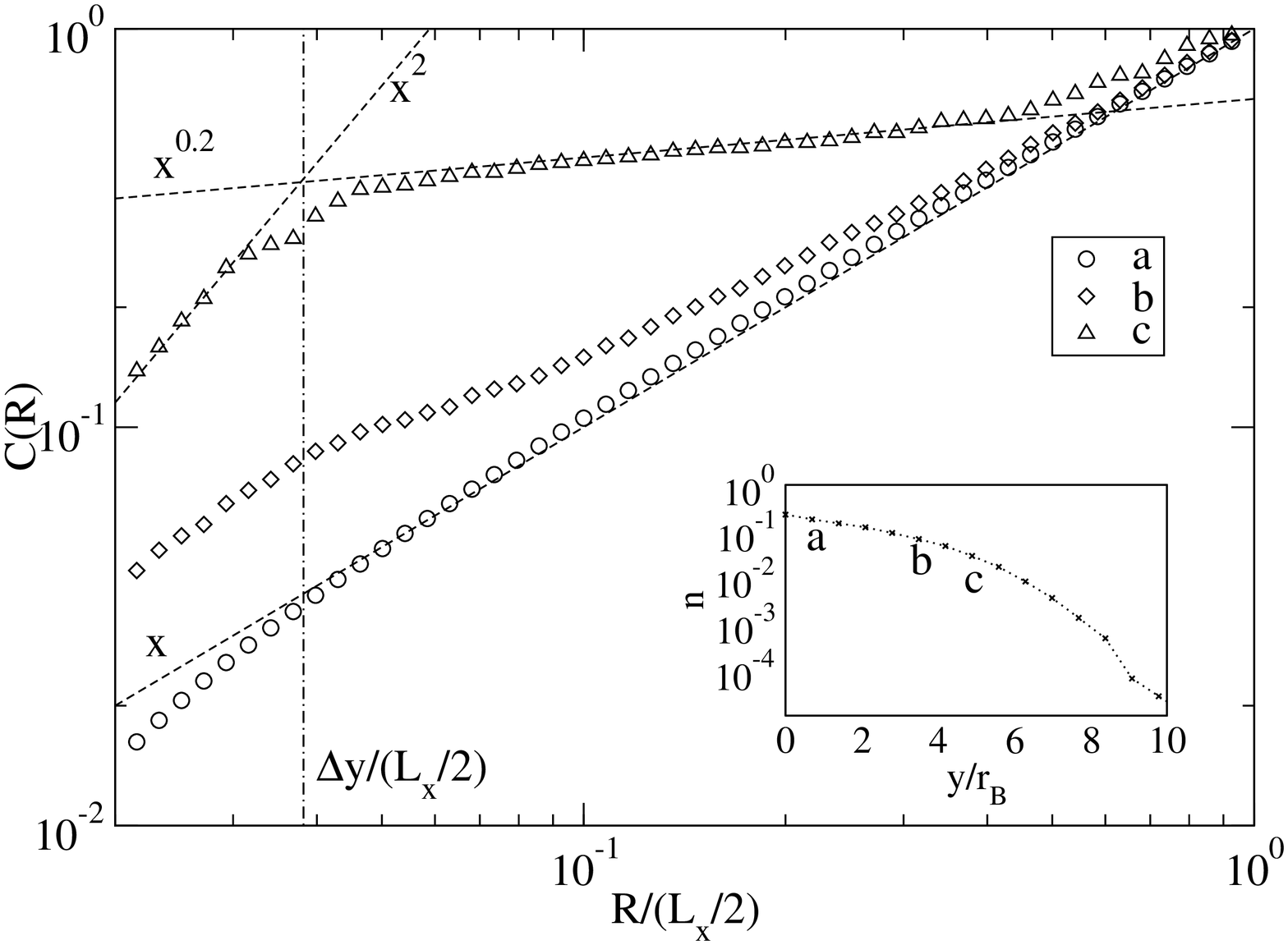}
\includegraphics[width=5.0cm,clip=true,angle=-90]{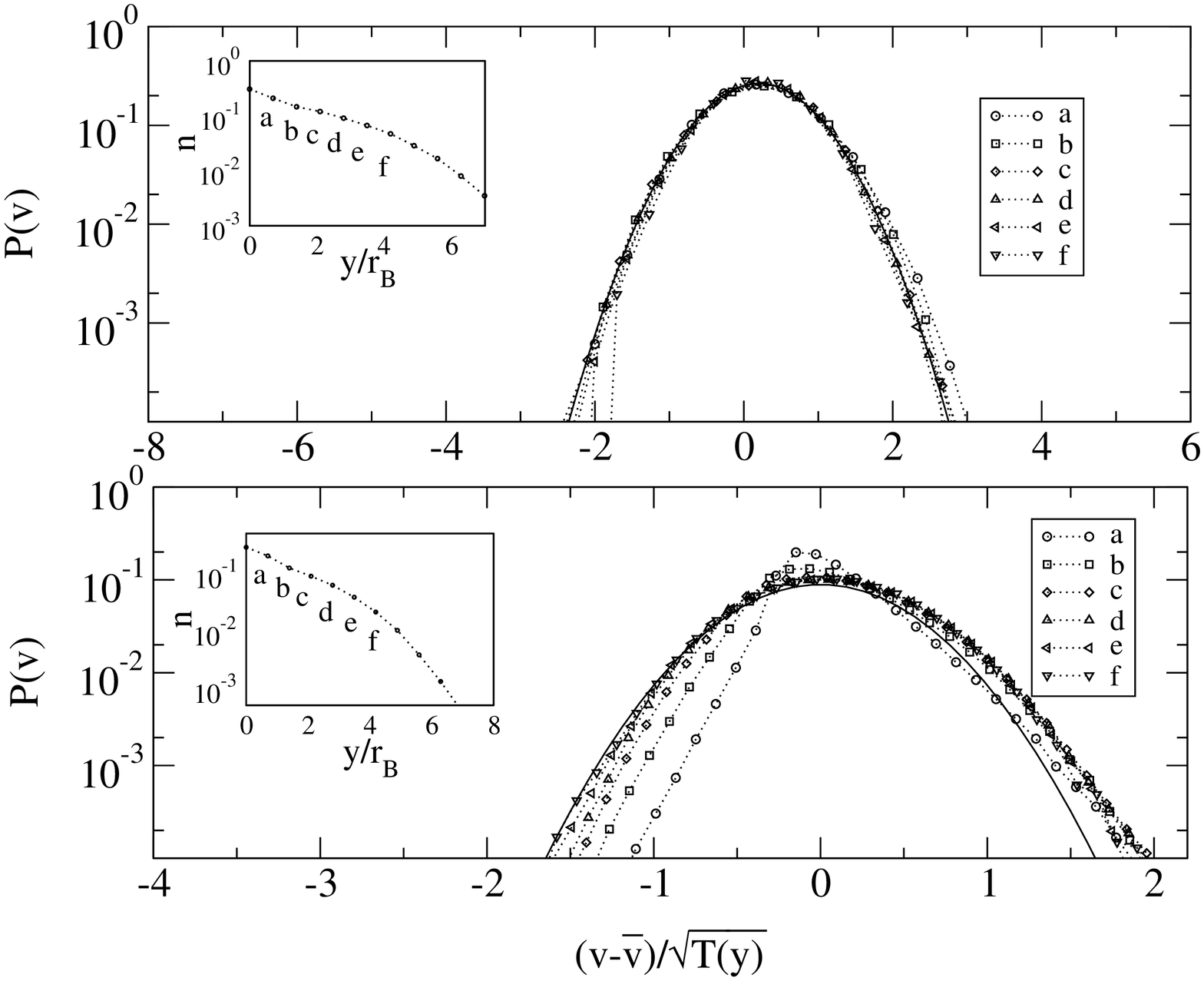}
\caption{Left: density-density correlation function
$C_{\Delta y}(y,R)$ in stripes
$a,b,c$ at different heights, as indicated in the inset,
for the non-homogeneous model with an inclined
bottom. The inset shows the density profile versus the rescaled height
$y/r_B$ and the letters $a,b,c$ locate the heights $y$ (or densities)
of the stripes $a,b,c$ chosen to compute $C_{\Delta y}(y,R)$.
Right: horizontal velocities pdf in stripes at different
heights for the same model. The inset is as above.}
\label{i2_C}
\end{center}
\end{figure}
Again, our discussion focuses on the density correlations 
$C_{\Delta y}(y,R)$ computed in stripes at different density 
(figure~\ref{i2_C}-left). 
Even in this system, clustering effects show up and are quantified 
by a correlation dimension $d_2$ ranging from $1$ 
in homogeneous stripes to $0.2$ for highly clustered stripes.  

The distribution of horizontal velocities in slabs at different heights
are plotted in the right frame of figure~\ref{i2_C}.  The emergence of
non-Gaussian behavior is clearly evident especially in the case with $r_w<r$
and mainly in the stripes near the bottom wall. Classical rheological model
proposed by Jenkins and Richman~\cite{jenkins} invokes a quasi-Gaussian
equilibrium to calculate the transport coefficients. The results of our
simulations, however suggest, that near the bottom wall, the Gaussian
approximation seems a very poor description of the real distribution.  This
is not only a consequence of inelasticity but also
an effect of the proximity to the boundary, where high spatial gradients can 
easily bring the gas out of equilibrium. 
More recent derivations of hydrodynamic equations~\cite{sela2, brey} 
use a Boltzmann-like approach
for inelastic gases which yields non-Gaussian velocity distributions: these
theories pose on a more solid basis and provide much more reliable 
estimations of transport coefficients.

\section{The problem of scale separation}
The reliability of hydrodynamics in the description of fluidized
granular gases has been intensively probed through simulations and 
experiments suggesting, in some cases, a certain range of validity
which surprisingly extends to very inelastic regimes.  However, even
in these lucky situations, the success is somehow lacking a rigorous
foundation and addresses the question ``why hydrodynamics works?''.
Goldhirsch~\cite{goldhirsch99} for instance pointed out that ``the
notion of a hydrodynamic, or macroscopic description of granular
materials is based on unsafe grounds and it requires further study''.
He argued that one of the main obstacle lies on the absence of a sharp
distinction between the spatio-temporal scales of the microscopic
dynamics and the relevant macroscopic scales.  The aim of this section
is to briefly review his arguments on this fundamental issue. We
remind that the validity of hydrodynamics and its correct derivation
is still a subject of debate as recent discussions testify~\cite{brey_tan}.

A standard granular experiment involves about $10^3 \pm 10^5$ grains and
a container with a linear size of few orders of magnitude larger than
the typical size of grains. Therefore the possibility to identify an
intermediate scale separating microscopic kinetics from macroscopic
hydrodynamics is rather doubtful.  The lack of scale separation is not
only a mere experimental limitation because in principle one can
imagine experiments involving an Avogadro's number of grains and very
large containers. Instead it is of conceptual nature and not only
related to granular materials but also to molecular gases when subject
to large shear rates or large thermal gradients.  In general when
velocity or the temperature fields vary significantly over a length of
a mean free path, no scale distinction occurs between microscopic and
macroscopic scales, accordingly the gas should be considered
mesoscopic.  In granular gases, this kind of {\em mesoscopicity} is
generic and not limited to the presence of strong forcing.  Moreover,
phenomena like clustering, collapse and avalanches typical of granular
dynamics strongly violate the {\em molecular chaos} condition required
by the Boltzmann's approach.  In mesoscopic systems, fluctuations are
expected to be larger by consequence the ensemble averages of observables
need not to be representative of their typical values.  Furthermore in
systems without a true scale separation, like turbulent fluids or
systems undergoing a second-order phase transition, one expects that
the constitutive relations relating fluxes to densities are scale
dependent.

The quantitative demonstration of the intrinsic mesoscopic nature of
granular gases stems from the equation
$T_g \propto \gamma^2 l_0^2/(1-r^2)$~\cite{goldhirsch93}, 
relating the local granular
temperature $T_g$ to the local shear rate $\gamma$ and to 
the mean free path $l_0$. 
The above relation holds until $\gamma$ can be considered a
slow varying (decaying) quantity with respect to much more rapid
damping rates of the temperature fluctuations.
Then, the ratio between the variation of the macroscopic velocity 
$\delta v \sim \gamma l_0$ due to the shear and the thermal speed 
$v_T = \sqrt{T}$ is proportional to $\sqrt{1-r^2}$.  
A part from very low values of $1-r^2$, the shear rate is
always large thus the Chapman-Enskog expansion leading to the 
hydrodynamic theory for the system should be generally carried out
beyond the Navier-Stokes order. 
The above consideration is a
direct consequence of the supersonic nature of granular
gases~\cite{goldhirsch99}. It is clear that a collision between two
particles moving in the same direction reduces their relative
velocity but not the sum of their momenta. In a number of such collisions, 
the velocity fluctuations may become very small with 
respect to $\delta v \sim \gamma l_0$. We have to say that  
even the notion of mean free path may become useless in a
shear experiment because the mean square particle velocity is given
by $\gamma^2y^2+T$ ($y$ being the direction of the shear).
When $y \gg \sqrt{T}/\gamma$ the distance covered by a
particle in the mean free time $\tau$ is
$l(y)=yl_0\gamma/\sqrt{T}=y\sqrt{1-r^2}$ that may
become much larger than the ``equilibrium'' mean free path $l_0$ and even
greater than the system size in the streamwise direction.
The ratio between the mean free time $\tau=l_0/\sqrt{T}$
and the macroscopic characteristic time of the problem $1/\gamma$,
reads again proportional to $\sqrt{1-r^2}$. Therefore a sharp
separation between microscopic and macroscopic time scales 
rigorously occurs only when $r \to 1$ independently of
system and grain sizes. Two serious problems thus arise:
a) the fast local
equilibration allowing to use equilibrium distributions as zeroth order
approximations is not obvious;
b) the stability studies based on the linearization of hydrodynamic equations 
become meaningless since they predicts instabilities
on time scales which hydrodynamics is not supposed to resolve.

Goldhirsch~\cite{goldhirsch99} has also shown that the absence of a neat
distinction in space/time scales implies a scale dependence of
fields and fluxes, in particular the pressure tensor
depends on the coarse graining resolution used to take local
averages. This is similar to what happens, for example, in turbulence,
where the ``eddy viscosity'' is scale dependent. Pursuing this analogy,
Goldhirsch has noted that an intermittent behavior can be observed in
the time series of experimental and numerical measures of the
pressure tensor. Single collisions, which are usually
averaged out in molecular systems, appear instead in granular systems
as ``intermittent events'' affecting the time behaviour of relevant 
observables. 
 
\section{Granular Temperature in a simple double-well model.}
  So far, through the review of some models of inelastic gases, we have given 
  evidences for the non-thermodynamics nature of the 
  parameter $T_g$ called ``granular temperature''. 
  We have indeed shown that, $T_g$ is usually
  different from the thermostat temperature, it can be very 
  inhomogeneous even in homogeneously driven systems and may strongly
  depend upon the scale of observation. Finally we have mentioned the fact 
  that in granular mixtures, $T_g$ does not govern the energy balance. 
  In this section we want to
  show that $T_g$ still maintains the role of parameter controlling the
characteristic times of the granular dynamics.
Here we discuss a simple toy model in
which the main ingredient of granular gases, the inelasticity, is at work, but
the dynamics is characterized by a time-scale determined by the granular
temperature through an Arrhenius-like formula~\cite{CeccoPRL}.

The model consists of two inelastic hard rods (the simplest granular gas)
constrained to move on a line under the effect of a bistable external
potential $U(x)=-ax^2/2+b x^4/4$. 
The system is coupled to a bath which exerts upon particles a 
velocity-dependent friction and a random force.  
In the absence of collisions, the particles evolve according to:
\begin{equation}
M \frac{d^2 x_i}{dt^2}=-M \gamma\frac{d x_i}{dt} - U'(x_i)+\xi_i(t)
\label{eq:kramers}
\end{equation}
where, prime indicates the spatial derivative, $x_i$ ($i=1,2)$ represents
the position of particles, $\gamma$ is a friction coefficient and 
$\xi_i(t)$ is the stochastic driving with variance $\propto T_b$.
 
\begin{figure}
\begin{center}
\includegraphics[clip=true,width=6cm,keepaspectratio]{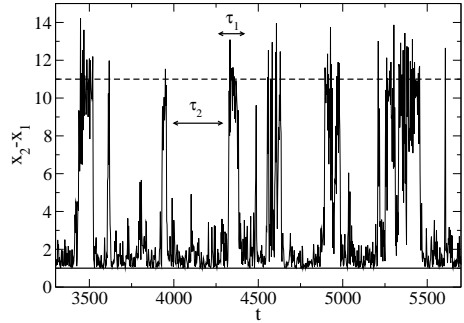}
\caption                                     
{Relative distance $x_2-x_1$ as a function of time for a system with
$r=0.9$ and $T_b=4.0$. The solid line indicates the diameter of the
rods $d =0.1$, while the dashed marks the well separation 
$L\simeq 10.95$, (for potential parameters $a = 0.5$ and $b=0.01$).}
\end{center}
\label{fig_xrel}
\end{figure}

\begin{figure}[htb]
\begin{center}
\includegraphics[clip=true,width=6.0cm,keepaspectratio]
{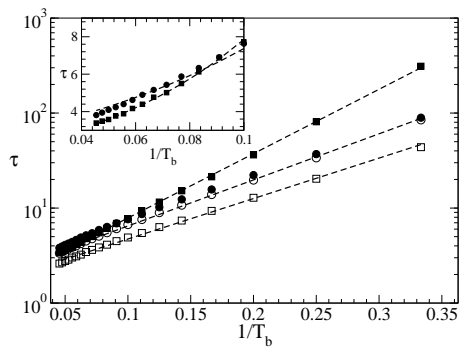}
\caption
{Arrhenius plot of mean escape times $\tau$. Open symbols refer to
elastic case: the escape time is $\tau_1$ (circles) when a well is
singly occupied, $\tau_2$ (squares) when a well is doubly occupied.
Full symbols, instead, correspond to the inelastic system ($r=0.9$).
Linear behavior indicates the validity of Kramers theory with
renormalized parameters and the slopes agree with values obtained from
Eq.~(\ref{newkramer}). Inset: enlargement of the crossover 
region where $\tau_2$ becomes smaller that $\tau_2$.  
The arguments of the exponentials (dashed lines) in the same
figure have been obtained by formula~(\ref{newkramer}).}
\label{fig_tempi}
\end{center}
\end{figure}

The basic phenomenology of the model is illustrated in Fig.~\ref{fig_xrel}.
The relative particle distance, $y=x_2-x_1$, fluctuates in time
showing time intervals of average lifetime
$\tau_2$, when particles are confined to the same well ($y\sim d$)
alternated with intervals, of average
lifetime $\tau_1$, when particle sojourn in separate wells.
The dynamics is dominated by two competing effects, the dissipation in
the collisions and the excluded volume. The first brings the particles
together in the same well (clustering) while the other favors their
staying apart in different wells.  
This two opposite effects are
responsible for the existence of $\tau_1$ and
$\tau_2$ as different time scales. 
Figure~\ref{fig_tempi} shows that $\tau_2$ and $\tau_1$ follow an 
Arrhenius behavior with a suitable parameter renormalization
with respect to the independent particle problem:
\begin{equation}
\label{newkramer}
\tau_k \approx \exp\bigg[\frac{W_k}{T_k}\bigg],
\end{equation}
where $k=(1,2)$ indicates single or double occupation of a well,
$W_1 = \Delta U$ ($\Delta U$ being the energy barrier between the wells)  
and $W_2 = \Delta U - \delta U<\Delta U$.  The correction $\delta U$
to the energy barrier $\Delta U$ amounts to 
$a(d/2)^2 + b/4(d/2)^4$ and takes into account the effect of the excluded
volume repulsion. When two grains belong to the same well their center
of mass lies higher than if they were in separate wells, then then each 
grain experiences a lower (effective) energy barrier. 
This is a typical correlation effect, because the repulsion makes 
less likely the double occupancy of a well, 
with respect to the non interacting case.
The smaller the ratio of the well width to the particle
diameter, the stronger the reduction of the escape
time~\cite{Tarazona,CeccoJCP2}.
In figure~\ref{fig_tempi} and related inset, the reader can see that the plots
of $\tau_1$ and $\tau_2$ of the inelastic system (dark symbols), intersect at
a certain temperature $T_b = T_c$. Below $T_c$ the time $\tau_2$ becomes
smaller than $\tau_1$.  The origin of this crossover lies on the fact that,
in the inelastic system, temperatures $T_2$ and $T_1$ are no longer
equal to $T_b$ and furthermore $T_2 < T_1$.  Thus, the mean lifetime of the
clustering regimes can be still described by
expression~(\ref{newkramer}), but now the granular effect competes
with the excluded volume correction, eventually leading to $\tau_1 > \tau_2$.

A simple argument can be used to estimate the
shift of $T_2$ from $T_b$.  For moderate driving intensity, $T_1$ is
nearly equal to $T_b$, while $T_2$ is lower than $T_b$ by a factor
which depends on the inelasticity. Simulations show that $T_2$ varies 
linearly with $T_b$ and its slope is a decreasing function of the
inelasticity $(1-r)$.
A good estimate of temperature $T_2$ can be obtained 
by an energy balance argument when the two particles belong to the 
same well regarded as an harmonic well $V(x)=\omega^2_{min} x^2/2$.
The average power per particle satisfies the balance equation   
\begin{equation}
\overline{\frac{dE}{dt}} = 2\gamma (T_b - T_2) + 
\overline{\frac{\delta E_c}{2\tau_c}} 
\label{eq:balance}
\end{equation}
where, $2\gamma (T_b - T_2)$ stems from the competition between the viscous 
damping ($-2\gamma T_2$) and the power supplied by the stochastic driving ($2\gamma T_b$).
While the last term in r.h.s of Eq.\eqref{eq:balance} estimates 
the mean power dissipated in each collision, $\tau_c$ being the typical
collision time.
From the rule \eqref{collision_rule} applied 
in 1-d, we have     
$\delta E_c = -(1-r^2)(v_2 - v_1)^2/4$.
At stationarity, we expect that $\overline{dE/dt} \sim 0$, thus 
$$
T_2 = T_b  - \frac{1-r^2}{8\gamma\tau_c}~\overline{(v_2-v_1)^2}.
$$ 
Assuming that the precollisional velocities $v_1$ and $v_2$ are nearly
independent we can approximate $\overline{(v_2-v_1)^2} \simeq \langle v_1^2 \rangle + \langle v_2^2 \rangle = 2 T_2$.

The collision time $\tau_c$, instead, is estimated through the oscillation
frequency in the harmonic well $\tau_c = \pi/\omega_{min}$ where
the factor $1/2$ stems from the excluded volume effect. Finally, we write
the formula
\begin{equation} 
\label{eq:Tg}
T_2  = \frac{T_b}{1 + q (1-r^2)}
\end{equation}
for the granular temperature, with $q = \omega_{min}/(4 \pi \gamma)$.
The knowledge of $T_2$ and $\Delta U$ characterizes the jump dynamics
of the system across the energy barrier even when the it is inelastic.  
Numerical simulations of Eqs.~\eqref{eq:kramers} verify very well
the relation~\eqref{eq:Tg}. 

This simple example demonstrates that the granular temperature, even
 if can not control the ``equilibrium'' behavior of a granular gas,
 fairly determines the typical dynamical times scales of the system.

\section{Conclusions}
We have summarized the main lines of a research carried out during
the last years on granular gases.
This paper focuses on the theoretical basis of a fluid-like description of
granular systems under strong external forcing. In experiments, the behavior 
of a granular gas strongly resembles that of a fluid. However it is never
at thermal equilibrium and, even though a kinetic temperature can be
defined and measured, it has not the same role of equilibrium temperature.
Moreover many conceptual concerns, such as the absence of space and time
scale separation, suggest that hydrodynamics is well posed only in a limited
range of parameters.  We have introduced a family of models of granular
gases under external forcing to investigate these issues.
Such models, addressing different physical situations, present common
features: strong correlated
density fluctuations (clustering), non-Gaussian behavior of velocity
distributions with heavy tails, and lack of energy equipartition or
thermalization.
The first model, an inelastic gas under external stochastic
driving, is of course the simplest and most idealized, but it displays
straightforwardly all these feature demonstrating that the main ingredient
leading to such anomalous behaviors is simply the inelasticity.

However the situation is not so hopeless: kinetic theories (used to build
hydrodynamics) 
work in the neighborhood of elastic limit, when all the above problems
appear in a mild form. More surprisingly, there are cases where some
predictions of usual statistical mechanics and thermodynamics are reliable 
also in strong inelastic conditions. We brought, as an example,  
the dynamics of a couple of granular particles in a double well potential  
which can be again characterized by an Arrhenius behaviour provided the 
environment temperature is replaced by the granular temperature. 
Furthermore, Some of
us~\cite{Puglisi:2002,Barrat:2004} have also showed that Green-Kubo relations
for response to linear perturbation are still valid, again substituting
granular temperature to external bath temperature (there have been several
attempts to derive Green-Kubo relations for granular gases, see for
example~\cite{gk}). Both these results are quite intriguing because appear to
be valid in strongly out-of-equilibrium regimes.


\end{document}